\documentclass{elsart3-1}

 \usepackage{graphics}
 \usepackage{graphicx}

\usepackage{amssymb}

\usepackage[english,francais]{babel}

\usepackage[usenames,dvipsnames]{color}

\usepackage{subfigure}

\newtheorem{e-proposition}[theorem]{Proposition}

\newtheorem{e-definition}[theorem]{Definition\rm}

\renewcommand{\theequation}{\arabic{equation}}
\def\Rey {{Re}}
\def\Uref {{U_0}}
\def\vel      {\mathbf{u}}

        \def\reffig#1{figure~\ref{fig:#1}}

\def\del#1{}

\begin{document}

\begin{frontmatter}


\selectlanguage{english}
\title{Travelling-wave solutions bifurcating from relative periodic orbits in plane Poiseuille flow}


\selectlanguage{english}
\author[authorlabel1]{Subhendu Rawat},
\author[authorlabel1]{Carlo Cossu} \&
\author[authorlabel2,authorlabel3]{Fran\c cois Rincon}

\address[authorlabel1]{Institut de M\'ecanique des Fluides de Toulouse, CNRS and Universit\'e de Toulouse\\ 
All\'ee du Professeur Camille Soula, 31400 Toulouse, France}
\address[authorlabel2]{Universit\'e de Toulouse; UPS-OMP; IRAP; Toulouse, France}
\address[authorlabel3]{CNRS; IRAP; 14 avenue Edouard Belin, 31400 Toulouse, France}


\medskip


\begin{abstract}
Travelling-wave solutions are shown to bifurcate from relative periodic orbits in plane Poiseuille flow at $\Rey=2000$ in a  saddle-node infinite period bifurcation.
These solutions consist in self-sustaining sinuous quasi-streamwise streaks and quasi- streamwise vortices located in the bulk of the flow.
The lower branch travelling-wave solutions evolve into spanwise localized states when the spanwise size $L_z$ of the domain  in which they are computed is increased.
On the contrary, upper branch of travelling-wave solutions develop multiple streaks when $L_z$ is  increased. 
Upper branch travelling-wave solutions can be continued into coherent solutions of the filtered equations used in large-eddy simulations  where they represent turbulent coherent large-scale motions.
%
%
%
}
\end{abstract}
\end{frontmatter}


\selectlanguage{english}

\section{Introduction}

The dynamics of transitional and turbulent wall-bounded shear flows is the subject of continued interest because of its numerous applications, ranging from drag-reduction for transport applications to the physics of planetary boundary layers.
A fruitful line of research has been to apply the dynamical systems approach to the understanding of subcritical transition in wall-bounded shear flows. 
In this context, the subcritical transition to turbulence has been related to the appearance of invariant solutions of the Navier-Stokes equations disconnected from the laminar basic state, and in particular of travelling-wave
  (TW) and relative periodic orbit (RPO) solutions.
Travelling-wave solutions, representing saddles in phase space, have been found at Reynolds numbers lower than the transitional ones in Couette flow, \cite{Nagata1990},
plane channel flow \cite{Ehrenstein1991,Waleffe2003} and pipe flow \cite{Faisst2003,Wedin2004}.
Relative periodic orbit solutions have also been computed in plane Couette flow \cite{Clever1997,Kawahara2001,Viswanath2007}, plane channel \cite{Toh2003} and pipe flows \cite{Duguet2008} and in the asymptotic suction boundary layer \cite{Kreilos2012}.
Global bifurcations of relative periodic orbits have been shown to be related to the transition to chaotic dynamics in plane Couette flow \cite{Kreilos2013} and in magnetohydrodynamic Keplerian shear flows \cite{Riols2013}.

Invariant solutions are thought to be important not only to understand the subcritical transition but also to gain understanding on the mechanisms of self-sustained turbulent motions at high Reynolds numbers. In particular,  it was shown in a recent study \cite{Rawat2015b} that the Nagata-Clever-Busse-Waleffe steady solution of the Navier-Stokes equations is connected to a solution of the filtered equations used in large-eddy simulations.
This suggests that this coherent steady solution is related to turbulent large-scale motions which are believed to be the dominant feature of the flow at high Reynolds numbers.

In a previous study \cite{Rawat2014} we have computed relative periodic orbit solutions of the Navier-Stokes equations in  a plane channel at Reynolds numbers ranging from $\Rey=2000$ to $\Rey=5000$ in a domain periodic in the streamwise and spanwise directions with extension $2 \pi h \times 2 h \times 2.416 h$, where $h$  is the channel half-width. 
These dimensions are  typical of large-scale motions in fully developed turbulent flows. 
The computed solutions occupy the bulk of the flow even at large Reynolds numbers  similarly to large-scale motions. 
Other solutions computed in the same flow at high Re do not have this feature and are most likely  relevant to the dynamics in the buffer-layer \cite{Itano2001,Toh2003} .

The computed relative periodic orbit solutions  \cite{Rawat2014} displayed the features of lower-branch solutions, which are relevant to the transition problem. 
In order to to track their `birth' trough a bifurcation and to compute the related upper branch solutions (which are expected to provide features more related to the developed turbulent flow), we attempted to continue these solutions to lower Reynolds numbers, but were unable to extend the continuation much below $Re = 2000$. 
In an alternative attempt, we therefore also tried to continue the solutions by extending the domain size. The main objective of this paper is to show that the periodic solution disappears in an infinite-period saddle-node bifurcation as the box spanwise size is increased, giving birth to  two branches of travelling-wave solutions. 
In the process, we find that the upper branch display the features of turbulent large-scale coherent motions and can be continued into a coherent solution of filtered large-eddy simulation equations, while  the lower-branch solution evolves into a localized edge-state when the spanwise size of the box is further increased.
A summary  of the problem formulation and techniques used to isolate periodic solutions in plane channel flow is given in \S2. The main results of the continuation are reported in \S3, and  discussed in \S4.

\section{Problem formulation and methods}

The pressure-gradient driven flow of an incompressible viscous fluid of constant density $\rho$ and kinematic viscosity $\nu$ in a plane channel of height $2\,h$ is considered. 
The flow satisfies the Navier-Stokes equations:
\begin{eqnarray}
\label{eq:NS1} 
\nabla \cdot \vel &=& 0, \\
\label{eq:NS2} 
\frac{\partial \vel}{\partial t} +  \vel \cdot \nabla \vel  &=&
-\nabla p + {1 \over \Rey} \nabla^2 \vel,
\end{eqnarray}
where the  Reynolds number Re=$\Uref h / \nu$ is defined with respect to the peak velocity $\Uref$ of the usual laminar (parabolic) Poiseuille solution. 
Dimensionless velocities have been defined with respect to $\Uref$, pressures with respect to $\rho \Uref^2$, lengths with respect to $h$ and  times with respect to $h/\Uref$. 
The streamwise axis aligned with the pressure gradient is denoted by $x$, while the wall-normal and spanwise coordinates are denoted by $y$ and $z$ and $u,v$ and $w$ are the velocity components along $x$, $y$ and $z$ respectively.
The flow is studied in the domain $[-L_x/2 , L_x/2] \times [-1,1]  \times [-L_z/2 , L_z/2]$. 
No-slip conditions are enforced in $y=\pm 1$ (walls) while periodic boundary conditions are enforced on the other boundaries.

In the final part of the study, new turbulent coherent solutions of plane Poiseuille flow are sought by solving the dynamical equations for filtered motions routinely used in large eddy simulations \cite{Hwang2010c,Hwang2011} .
The equations for the filtered motions are the usual ones (see e.g. \cite{Deardorff1970,Sagaut2006}):
\begin{equation}
\label{eq:LES}
 \frac{\partial \overline{u}_{i}}{\partial x_i} = 0;~~~~~ 
 \frac{\partial  \overline{u}_{i}}{\partial t} + \overline{u}_{j} \frac{\partial \overline{u}{_i}}{\partial x_{j}} = -\frac{\partial \overline{q}} {\partial x_{i}} + \nu \frac{\partial^2 \overline{u}_{i}} {\partial x^2_{j}}-\frac{\partial \overline{\tau}^r_{ij}}{\partial x_{j}},
 \end {equation}
where filtered quantities are denoted by an over-bar and 
$\overline{\mathbf \tau}^r = \overline{\mathbf \tau}^R - tr(\overline{\mathbf \tau}^R)\, \mathbf{I} /3$,
with
$\overline{\tau}^R_{ij}=\overline{u_i u_j}-\overline{u}_i \overline{u}_j$
and $\overline{q}=\overline{p}+ tr(\overline{\mathbf \tau}^R) / 3$.
We use Smagorinsky \cite{Smagorinsky1963} subgrid model based on the eddy viscosity $\nu_t$ for the anisotropic residual stress tensor $\overline{\tau}_{ij}$:
$\overline{\tau}_{ij}^r =-2\nu_t \overline{S}_{ij}$,
where $\overline{S}_{ij}$ is the rate of strain tensor associated with the filtered velocity field, 
 $\nu_t =D (C_s \overline{\Delta})^2\overline{\mathcal{S}}$ and 
$\overline{\mathcal{S}} \equiv (2\overline{S}_{ij}\overline{S}_{ij})^{1/2}$, 
$\overline{\Delta}=(\overline{\Delta}_x\overline{\Delta}_y\overline{\Delta}_z)^{1/3}$.
The Smagorinsky constant reference value is set to $C_s=0.05$, as in \cite{Hwang2010b,Hwang2011}, which is known  to provide the best performance for {\it a posteriori} tests \cite{Hartel1998}.
To avoid non-zero residual velocity and shear stress at the wall we use the wall (damping) function 
$D=1-e^{-(y^+/A^+)^2}$ 
with $A^+=25$.

Travelling-wave solutions are computed with a Newton-based iterative method implemented in the code {\tt peanuts}  \cite{Herault2011,Riols2013} which includes parameter continuation and is also used to analyze the linear stability of the converged solutions.
{\tt Peanuts} relies on repeated calls to numerical time-integrations of the Navier-Stokes or of the filtered equations equations to perform Newton iterations using a matrix-free iterative method.
As in many of previous similar investigations (e.g. \cite{Waleffe2003,Rawat2014,Kreilos2012}), the search space of the solutions is restricted to those with mid-plane reflection symmetry $\{u,v,w\}(x,y,z)=\{u,-v,w\}(x, -y, z)$, which reduces the number of degrees of freedom and is known to improve convergence.

The Navier-Stokes simulations are performed with  {\tt channelflow} which is based on a Fourier-Chebyshev-Fourier spatial pseudo-spectral discretization  \cite{Gibson2008}. 
Solutions are advanced in time using a second order Crank-Nicolson Runge-Kutta time stepping. 
Converged solutions were obtained with $32 \times 65 \times 32$ points in the streamwise, wall-normal and spanwise directions and enforcing a constant volume flux during the simulation.  
These solutions are almost identical to those computed on a coarser grid $16 \times 41 \times 16$.
The numerical results  were further tested by recomputing the same periodic solutions on the same grid with the {\tt diablo} code \cite{Bewley2008}.
{\tt diablo} was also used to perform the time-integrations of the filtered equations (LES) as in previous related investigations \cite{Hwang2010c,Hwang2011,Rawat2015b}.

\section{Results}

\subsection{From relative periodic orbits to travelling waves}

\begin{figure}
  \centering
   \centerline{
  \subfigure{\includegraphics[width=0.45\columnwidth]{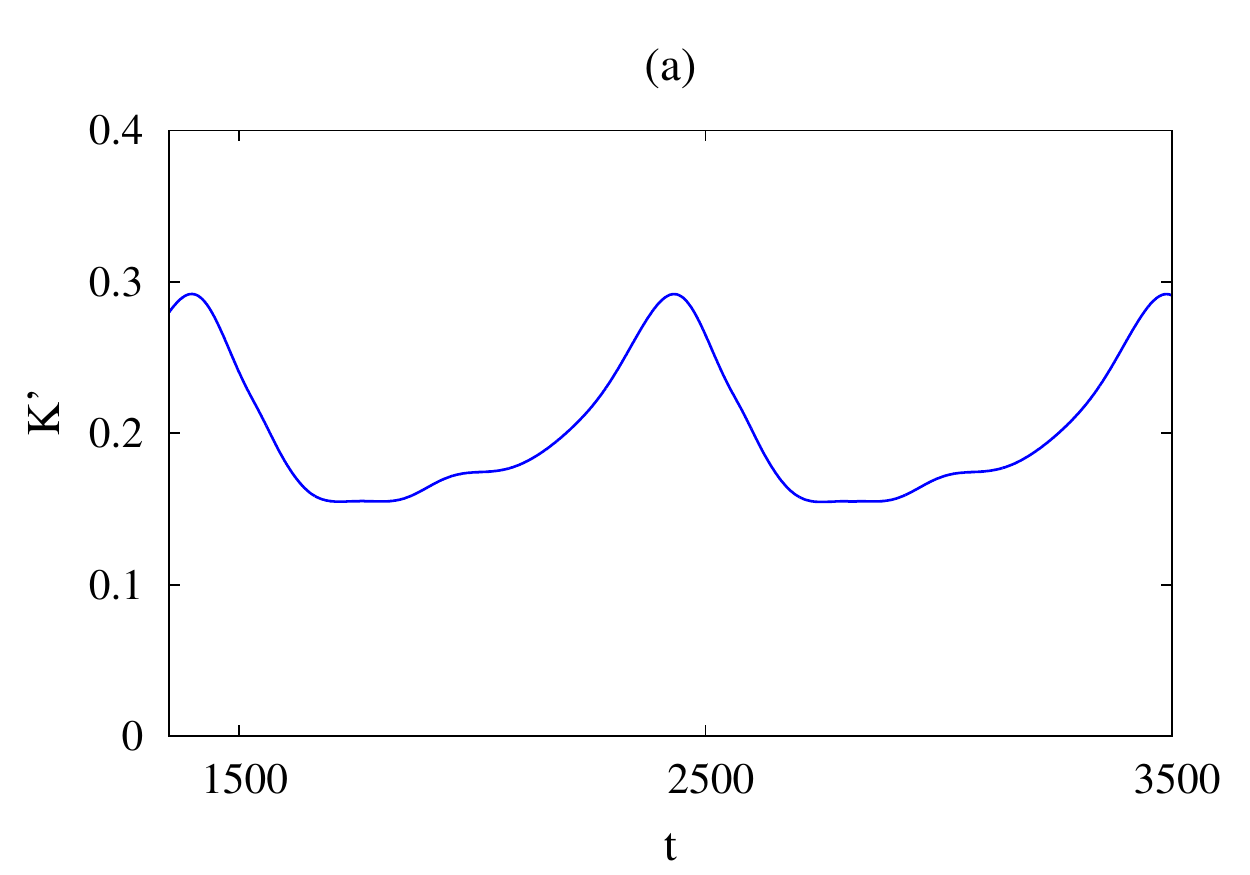}}
  \subfigure{\includegraphics[width=0.45\columnwidth]{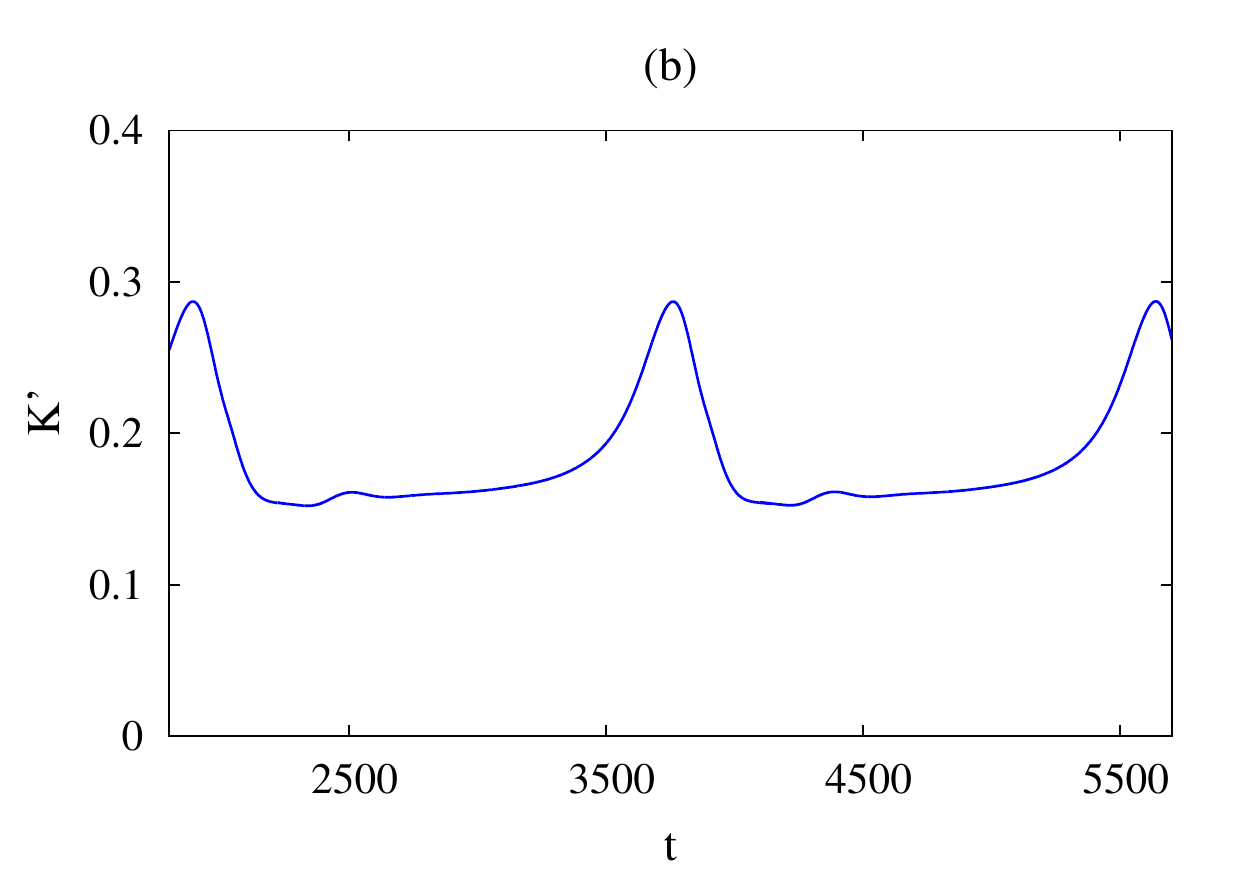}}%
   }
  \hspace*{\fill}%
  \subfigure{\includegraphics[width=0.45\columnwidth]{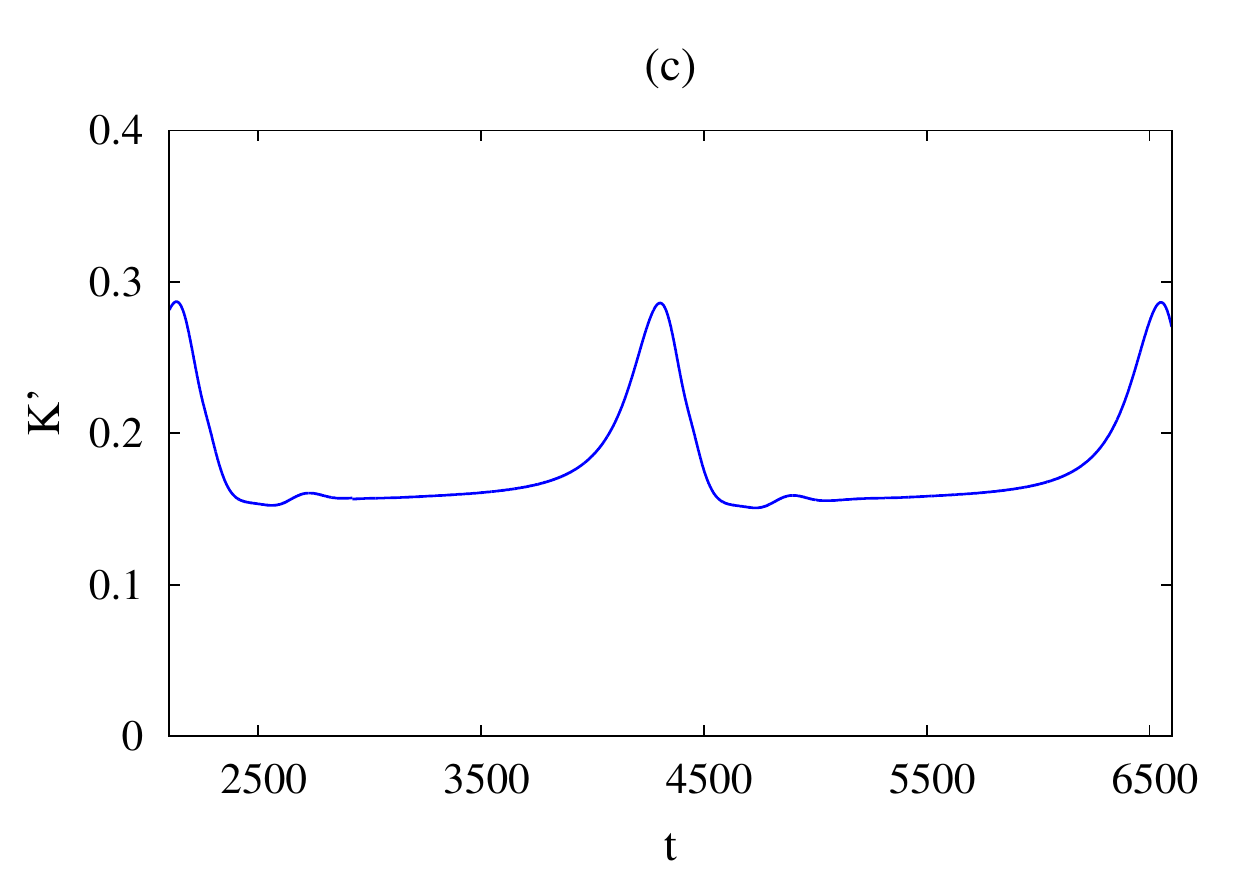}}
  \subfigure{\includegraphics[width=0.45\columnwidth]{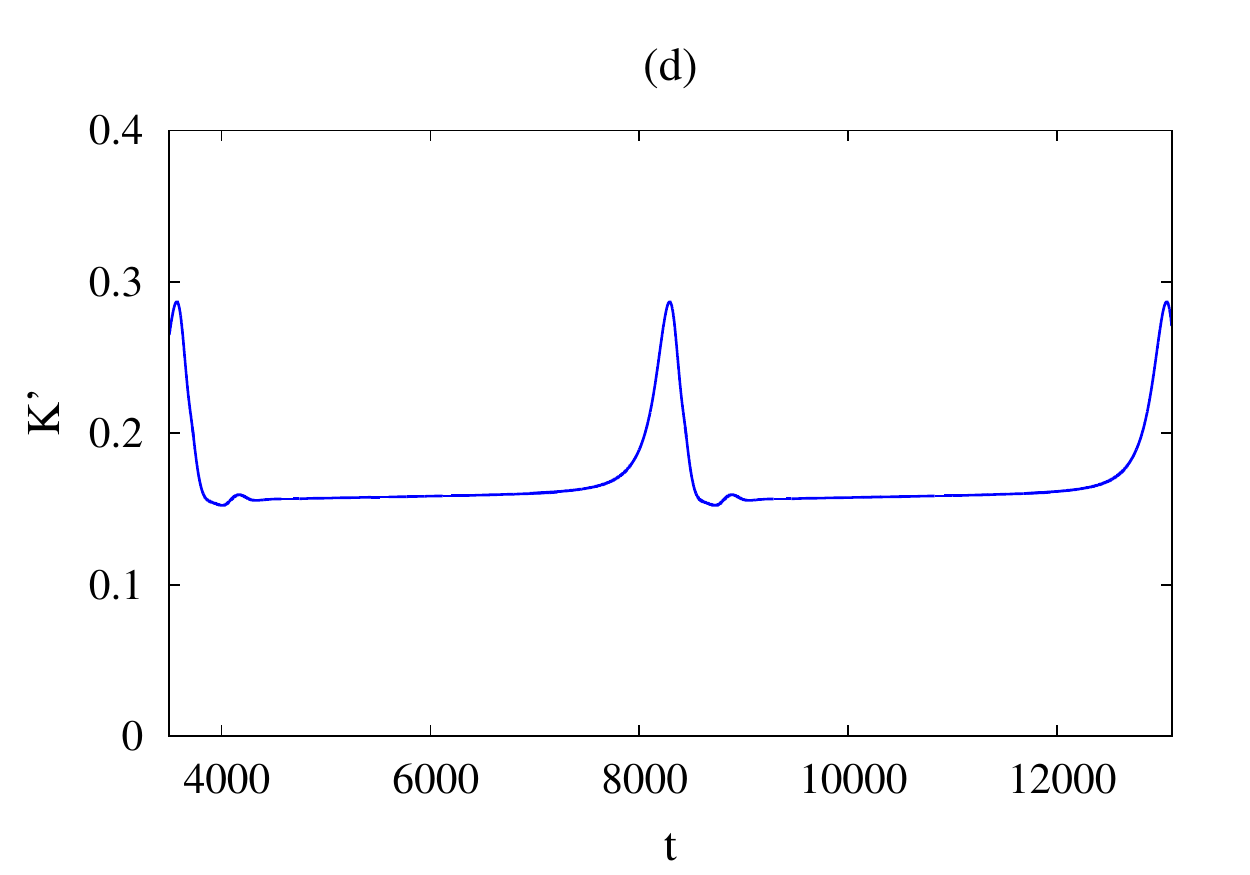}}%
  \hspace*{\fill}
  \caption{\small  Perturbation energy associated to the relative periodic orbit solutions computed with the edge-tracking procedure for the selected increasing values of the spanwise size $L_z= 3.525$ ,  $L_z= 3.535$ , $L_z= 3.540$ , $L_z= 3.548$ (panels $a$ to $d$). 
The initial transient before the convergence to the edge state is discarded.
Remark the increasing period of the solutions as $L_z$ is increased.}
  \label{fig:SNIPER_TimeHist}
\end{figure}
\begin{figure}
\centering
   \centerline{
 \includegraphics[width=0.5\columnwidth]{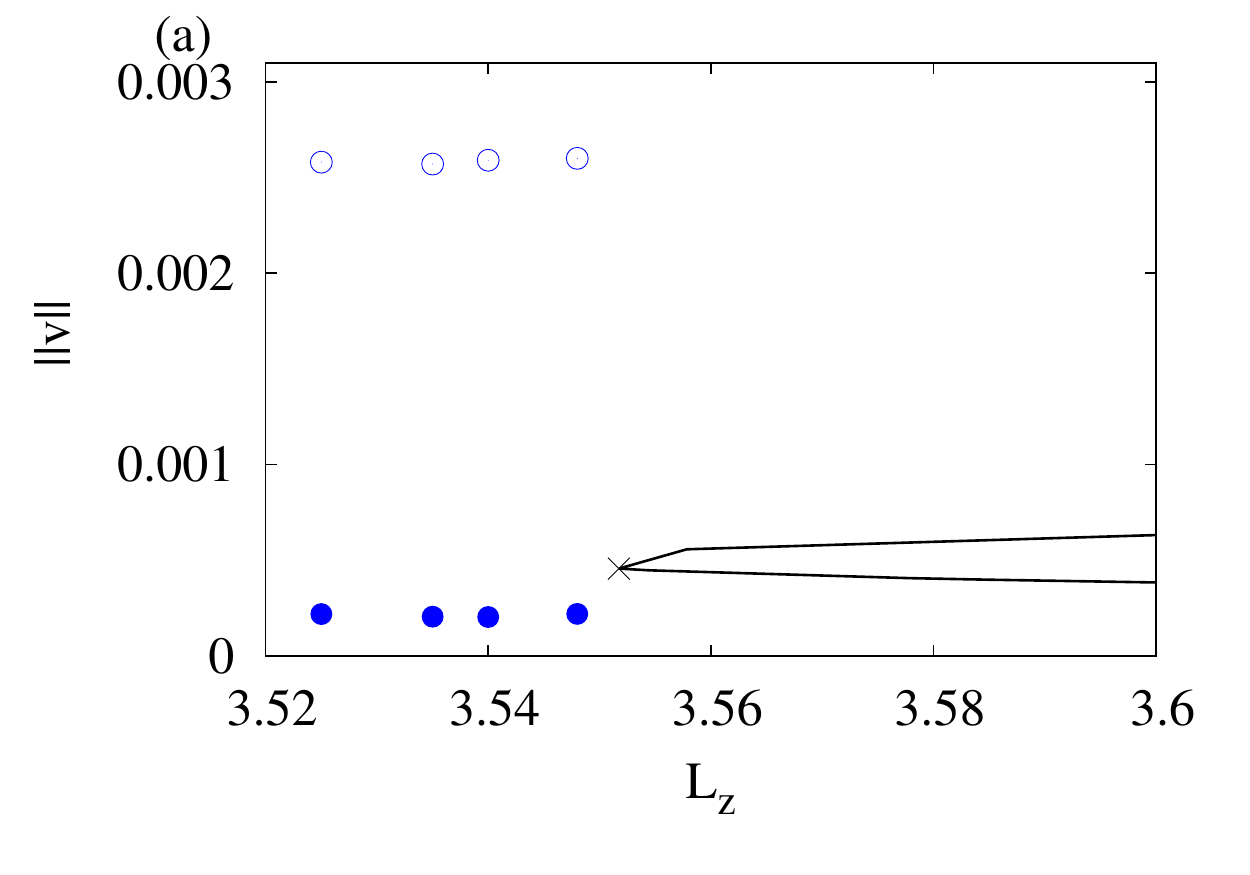}
 \includegraphics[width=0.5\columnwidth] {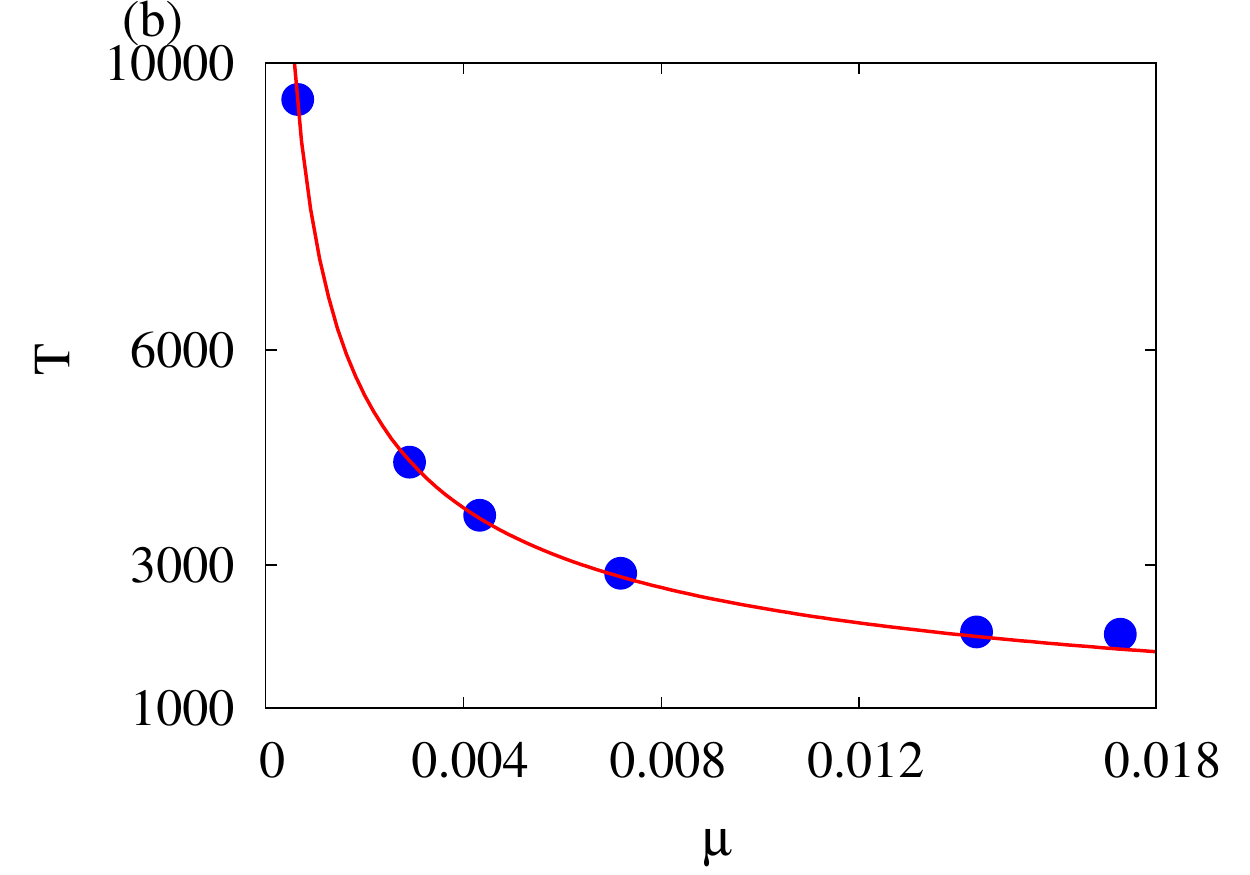}
}
  \vspace*{-1mm}
  \caption{\small (a) Bifurcation diagram in $L_z$, with $L_x = 2 \pi$ and $\Rey=2000$ where the relative periodic orbit solution disappears in a global saddle-node infinite period bifurcation originating a pair of travelling-wave solutions. 
The periodic solution is reported with its maximum (empty symbols) and minimum (filled symbols) values of the spatially-averaged wall-normal velocity. 
The symbols correspond to the $L_z$ values considered in \reffig{SNIPER_TimeHist}.
The point where the upper and lower branch travelling-wave solutions are generated is denoted by the $X$ symbol.
(b) Variation of time period $T$  close to the bifurcation point. 
The function  $T \approx 240/\sqrt\mu$ (solid line), where $\mu = (L_z^{*} - L_z)/L_z^{*}$,  fits well the data (symbols).  The four leftmost symbols correspond to the time-series reported in \reffig{SNIPER_TimeHist}.
}
\label{fig:LzCONT}
\end{figure}
\begin{figure}
\centering
   \centerline{
 \includegraphics[width=0.65\columnwidth]{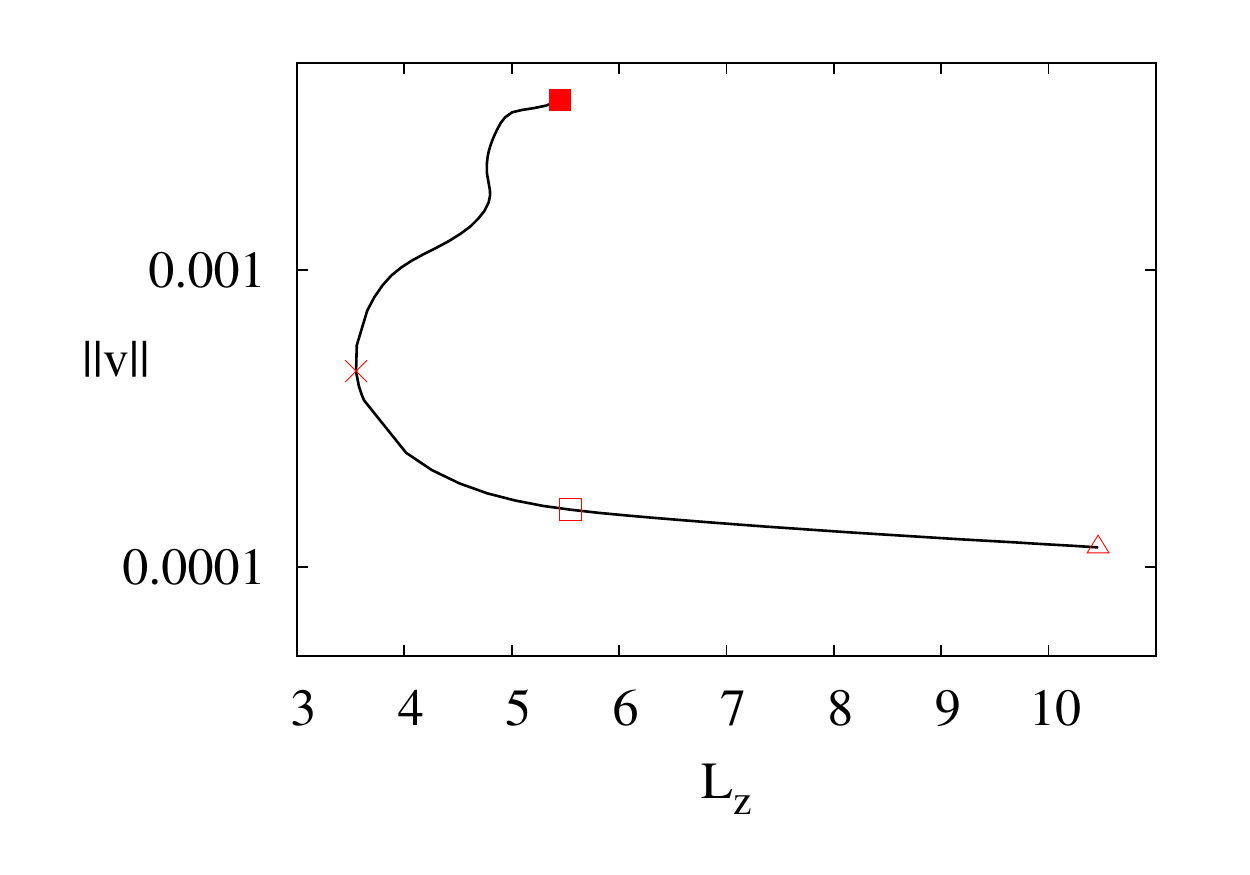}
}
  \vspace*{-1mm}
  \caption{\small Continuation diagram for travelling-wave solutions as a function of  $L_z$, with $L_x = 2 \pi$ and $\Rey=2000$. The left-most part of the diagram is already reported in  \reffig{LzCONT}(a).
The turning point where the upper and lower branch of the travelling-wave solution originate corresponds to the saddle-node infinite period global bifurcation. }
\label{fig:LzCONT_NTW}
\end{figure}

In a previous investigation \cite{Rawat2014},  relative periodic solutions were computed in plane Poiseuille flow in the domain of extension $2 \pi \times 2 \times 2.416$, for which Waleffe's travelling-wave solutions appear at the lowest Reynolds number \cite{Waleffe2003}, and for Reynolds numbers ranging from $\Rey=2000$ to $5000$. 
The period $T$ of these solutions increased with Reynolds number ($T=368$ for $\Rey=2000$, $T=739$ for $\Rey=3000$, $T=1090$ for $\Rey=4000$ and $T=1418$ for $\Rey=5000$) and the solutions were found to travel in the streamwise direction with a phase speed $C_x \approx 0.98$.
We have not found it possible to continue these  periodic  solutions  much below $\Rey = 2000$ leaving obscure their origin.  
Previous investigations of plane Couette flow, however, showed that the Nagata-Clever-Busse-Waleffe branch of steady solutions 
can be connected to a branch of periodic solutions by changing the box size \cite{Kawahara2001}.
We have therefore continued the relative periodic orbit solution obtained at $L_z=2.416$ to higher values of $L_z$ by using edge-tracking in the $\pm y$ symmetric subspace while keeping constant the Reynolds number to $Re =2000$ and the streamwise box dimension  to $L_x=2 \pi$. 

When $L_z$ is increased the periodic solutions display increasingly long quiescent phases separated by relatively quick bursts, as shown in \reffig{SNIPER_TimeHist}.
This leads to an increase of the period $T$ of the solutions with $L_z$.
For values above $L_z \approx 3.55$ the edge tracking converges  to a travelling wave (TW) solution instead of a relative periodic orbit.
The continuation of the TW solution in $L_z$ using pseudo-arclength continuation  based on the Newton-based iterations of {\tt peanuts} reveals the existence of an upper and a lower-branch of solutions connected by a saddle node bifurcation at $L_z^{*} = 3.55$ as shown in  \reffig{LzCONT}$(a)$ and in \reffig{LzCONT_NTW}.

The bifurcation observed at $L_z^{*}$  is global and is associated to a divergence of the period $T$ of the periodic solutions while approaching the bifurcation.
As shown in \reffig{LzCONT}$(b)$, the data are well matched by the fit $T = 240/{\sqrt{\mu}}$,  where $\mu = (L_z^{*} - L_z)/L_z^{*}$.
The divergence of $T$ as $\mu^{-1/2}$ and the \textit{O(1)} amplitude of the periodic solution while approaching the critical value $L_z^*$ are the hallmark of the  saddle-node infinite-period  bifurcation where a periodic solution is replaced by a pair of fixed points (see e.g. \cite{Strogatz2001}). 
The same type of global bifurcation has already been found e.g. in the study of axisymmetric convection \cite{Tuckerman1988} and, more recently, in the homotopy of  plane Couette flow travelling-wave solutions which are continued to asymptotic suction boundary layer (ASBL) solutions by increasing the suction velocity on the lower wall  \cite{Kreilos2013}. 
In all these cases, including ours, an upper and a lower branch travelling-wave solutions and their spanwise shifted heteroclinically connected copy collide in phase space at $L_z^{*}$, in a pair of saddle-node bifurcations which result in the formation of a periodic solution which repeats itself with a shift symmetry every $T/2$.

\subsection{Structure of  travelling-wave solutions}

\begin{figure}[htp]
  \centering
   \centerline{
 \includegraphics[height=0.28\columnwidth]{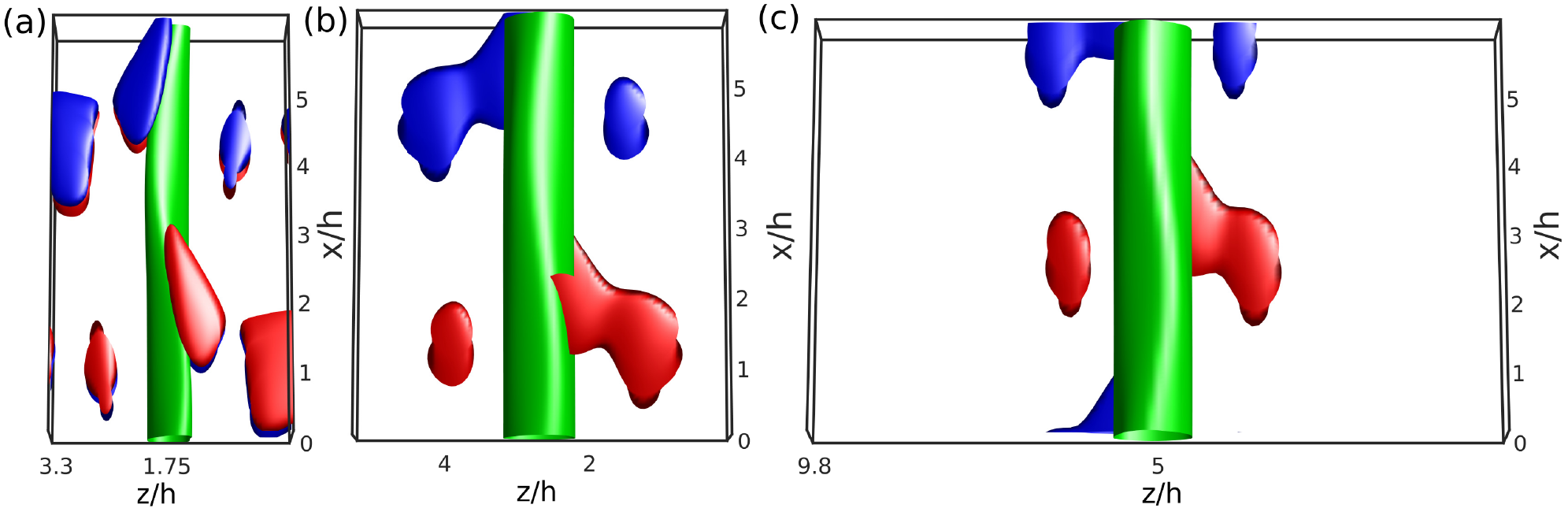}
  }

  \caption{\small Flow fields associated to the travelling-wave solutions corresponding to the:
(a) saddle node $L_z \approx 3.55$ (X symbol in \reffig{LzCONT_NTW}), while
(b) and (c) are the lower branch solution respectively computed for $L_z = 5.55$ (empty square symbol in \reffig{LzCONT_NTW}) and $L_z=10.5$ (empty triangle symbol in \reffig{LzCONT_NTW}).
The iso-surfaces at $u^+= -2$ are plotted in green, while red and blue surfaces correspond to  positive and negative streamwise vorticity at $\omega_x = \pm 0.65 max(\omega_x)$
\label{fig:LZCONT_LB}
}
\end{figure}
\begin{figure}
\setlength{\unitlength}{1cm}
\begin{center}
\includegraphics[width=0.45 \columnwidth]{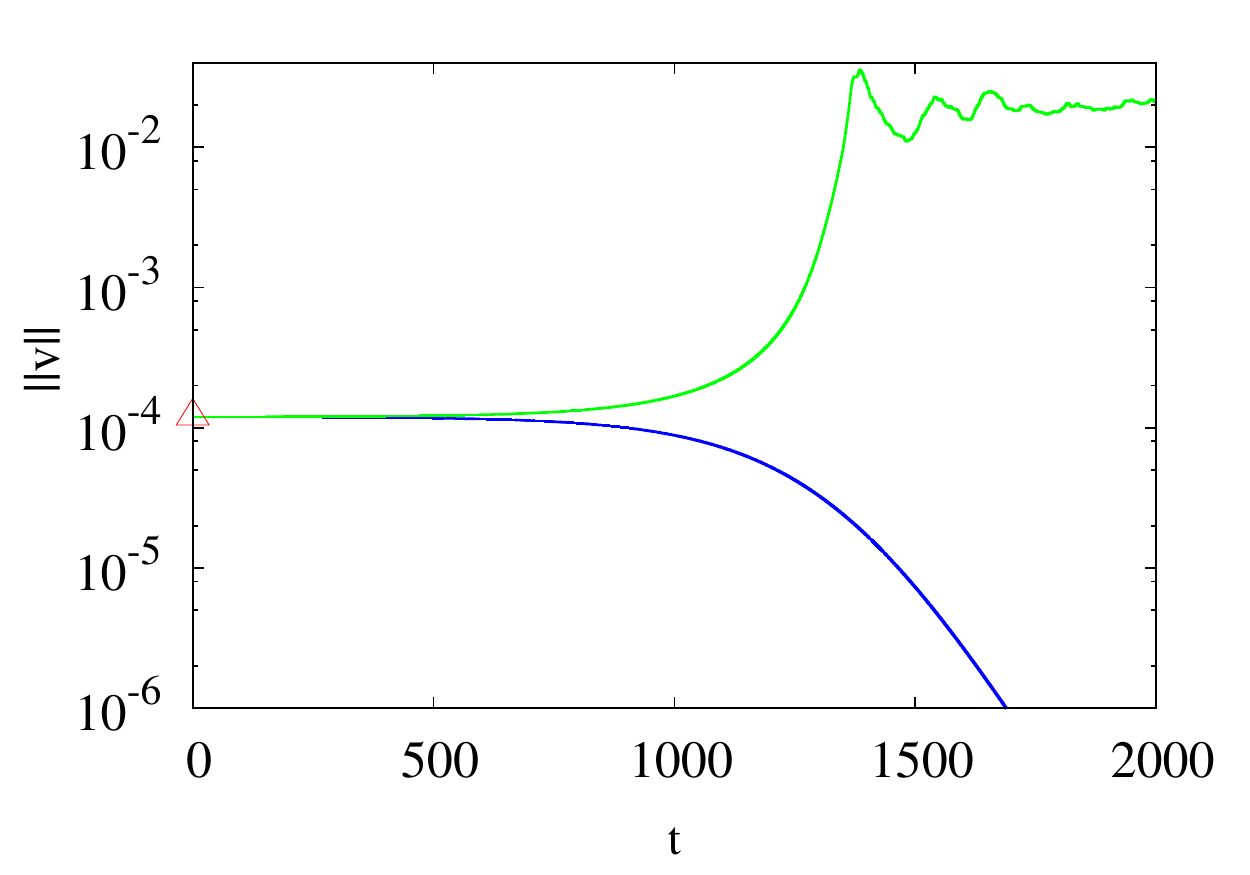} 
\includegraphics[width=0.45 \columnwidth]{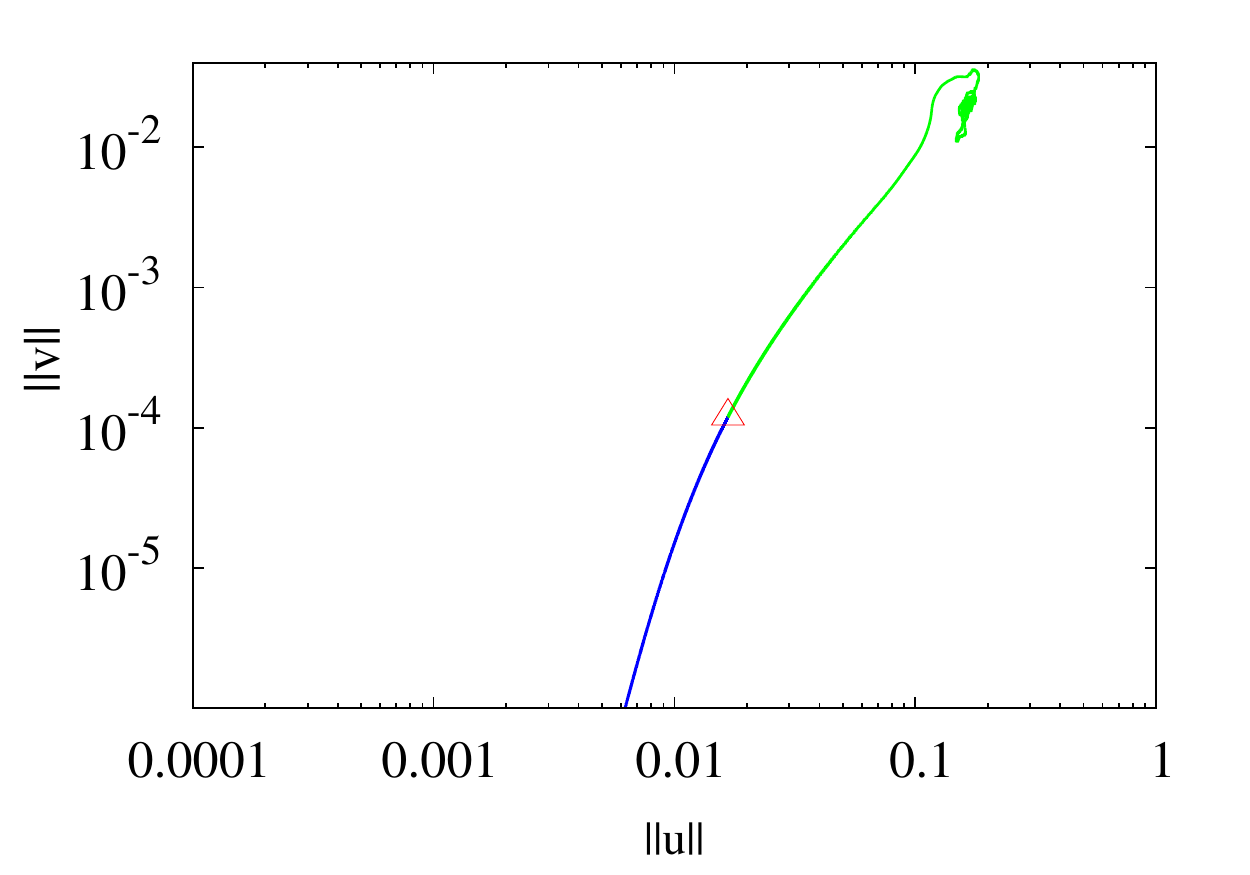} 
\end{center}
\caption{\small Trajectories initialized along the unstable eigendirection of the lower branch spanwise localized travelling-wave solutions ($L_x=2 \pi$, $L_z=10.5$, $\Rey=2000$) and represented in the $t - \|u'\|$ (left panel)  and $ \|u'\| - \|v'\|$ (right panel) planes respectively.
When initialized along one direction of the unstable manifold the flow rapidly relaxes to the laminar Poiseuille solution (blue line). When initialized in the opposite direction, a turbulent state is attained (green line). This indicates that the lower branch localized state (red cross in the figure) sits on the edge of chaos.
}
\label{fig:EDGE_LB_LOC}
\end{figure}

The flow structures associated to travelling-wave solution at the bifurcation consists of the usual sinuously bent low-speed streak flanked by a pair of quasi-streamwise vortices, as shown in \reffig{LZCONT_LB}$(a)$.
When this solution is continued to higher values of  $L_z$ along the lower branch, at least up to  $L_z = 10.5$  the streaky structure  remains unique and becomes localized in the spanwise direction, leaving an increasingly large portion of the domain almost unperturbed  as shown in \reffig{LZCONT_LB}$b$ and $c$) 
These localized structures, which arise `naturally' through continuation without the need to use a windowing function as e.g. in  \cite{Gibson2014}, are unstable. 
For $L_z=10.5$ the unique unstable eigenvalue found is $5.79 \times 10^{-3}$  and is therefore an edge state as can be verified  by perturbing the solution along its unstable manifold, i.e. with the unstable eigenfunction.
It is indeed found  that this perturbation leads to a turbulent state or to a fast decay to the laminar state depending on the sign of the perturbation,  as  shown in \reffig{EDGE_LB_LOC}.
The observed spanwise localization of the lower-branch solution is similar to that observed in a number of other flows  \cite{Duguet2009,Duguet2012,Khapko2013,Gibson2014}, and therefore seems to be a generic property of lower branch solutions in shear flows.
\begin{figure}
\begin{center}
\includegraphics[width=0.3\columnwidth]{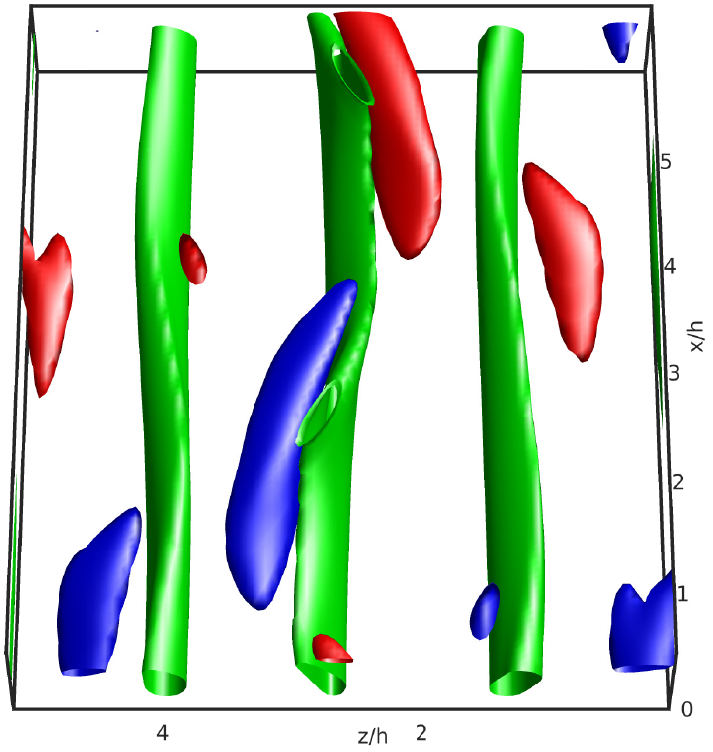}
\end{center}
  \caption{\small  Flow field associated to the upper branch solution computed at $L_z = 5.55$ corresponding to the filled square symbol in \reffig{LzCONT_NTW}.
The iso-surfaces at $u^+= -2$ are plotted in green, while red and blue surfaces correspond to  positive and negative streamwise vorticity at $\omega_x = \pm 0.65\,max(\omega_x)$
\label{fig:LZCONT_UBLB}
}
\end{figure}

\begin{figure}
 \centering
 \includegraphics[width=0.32 \columnwidth]{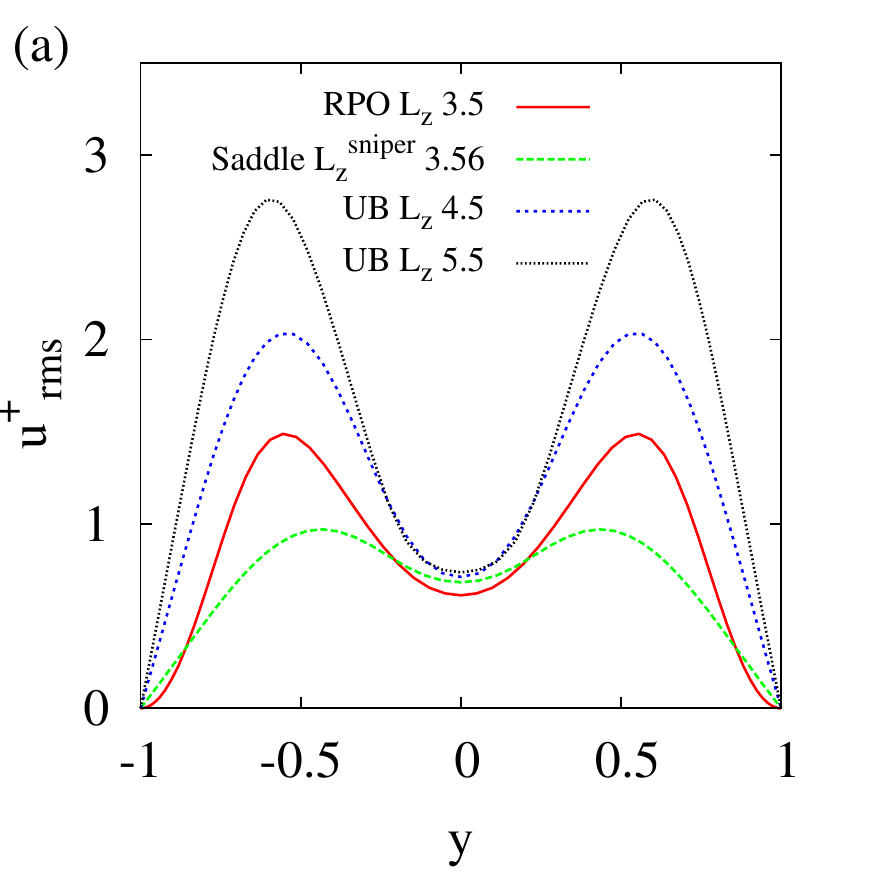}
\includegraphics[width=0.32 \columnwidth]{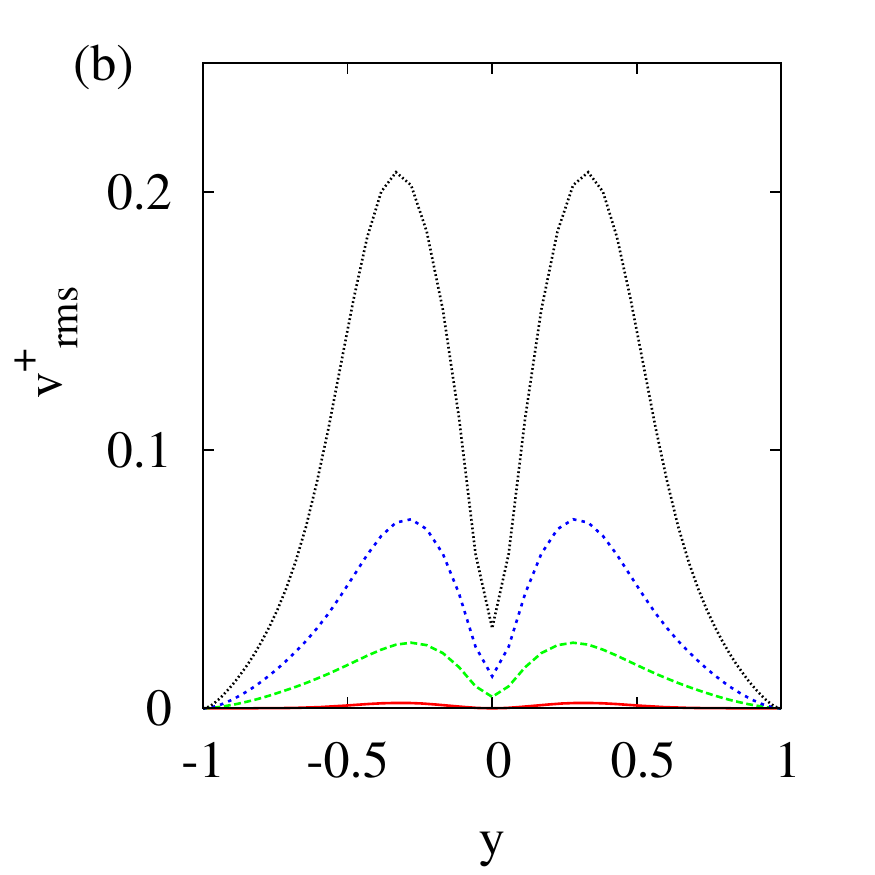}
\includegraphics[width=0.32 \columnwidth]{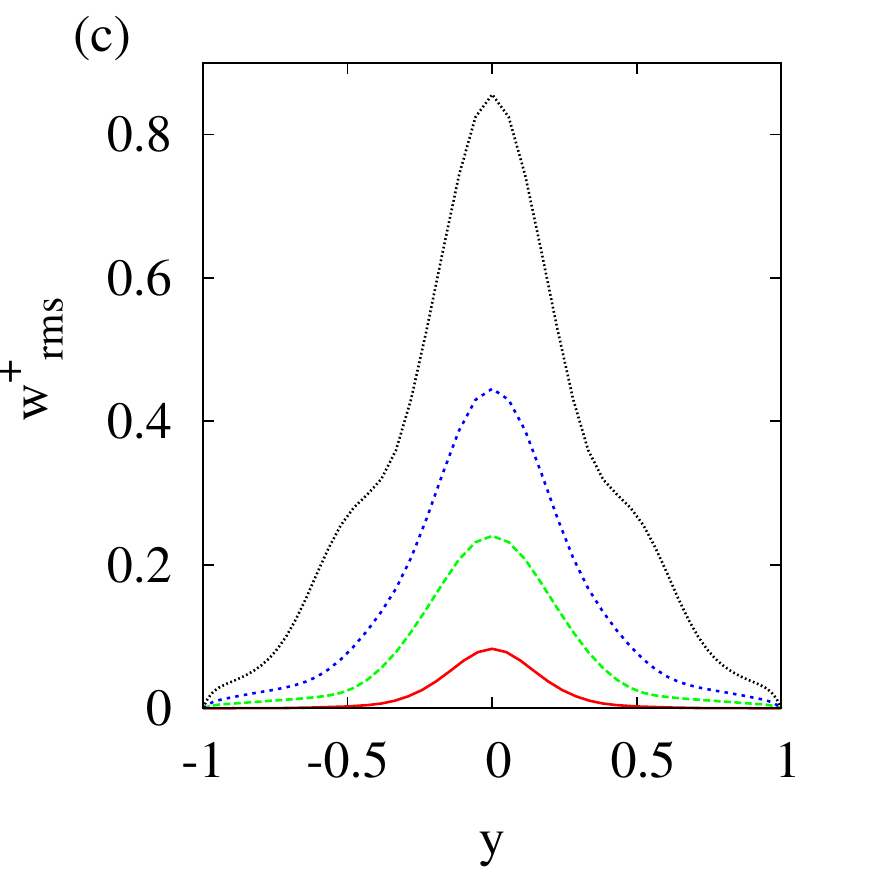}
 \caption{\small Comparison of the $rms$ velocity profiles of the travelling-wave upper branch solutions for selected values of $L_z$ to the $rmr$ profiles of the relative periodic orbit solutions existing for $L_z < L_z^*$.}
 \label{fig:LZCONT_COMPSOLS}
\end{figure}

The evolution of the upper-branch travelling-wave solution is completely different.
In this case, indeed, continuation to higher values of  $L_z$ leads to an increase in the number of streaky structures. 
For instance, at $L_z=5.55$, as reported in \reffig{LZCONT_UBLB}(b),  the upper branch solution contains three low speed streaks and three pairs of quasi-streamwise vortices which correspond to a streak spacing $\lambda_z \approx 1.8$,  in good agreement with the size of large-scale motions (LSM) in the turbulent channel \cite{delAlamo2003,Hwang2010b}.
This spacing is confirmed by the analysis of the spanwise premultiplied spectrum of the streamwise velocity (not shown).
The energy of the upper branch solutions increases for increasing $L_z$.
In figure~\ref{fig:LZCONT_COMPSOLS} it can be seen that during the continuation the $rms$  ($x-z$ averaged) profiles of the three velocity components preserve a qualitatively similar shape, which is also similar to that of the relative periodic solutions existing before the global bifurcation and this despite the changing nature of the underlying solutions.

Despite the changing nature of these solutions
It has been impossible to continue the upper branch to values larger than $L_z \approx 5.57$.
The relevance of the upper-branch solution to the dynamics of turbulent large-scale coherent structures is further investigated using filtered large-eddy simulations to include the locally averaged effect of small-scale motions, as discussed in the next section.

\subsection{Continuation from Navier-Stokes solutions to coherent solutions of the filtered equations}

In recent investigations \cite{Rawat2015b} it was shown that  invariant solutions of the Navier-Stokes equations can be continued to invariant solutions of the equations for filtered turbulent motions used in large-eddy simulations.
In the equations used in large eddy simulaitons, small-scale motions are averaged by filtering and modelled by a sub-grid model.  
In these previous studies the Smagorinky's 1963 model \cite{Smagorinsky1963} was used  along the lines of previous investigations of the self-sustained processes at large scale in turbulent shear flows \cite{Hwang2010b,Hwang2011}. 
In this context the Smagorinsky constant $C_s$ is used as a continuation parameter. 
The value $C_s=0.05$ corresponds to large eddy simulations having good a posteriori agreement with direct numerical simulations \cite{Hartel1998}, while Navier-Stokes solutions are obtained with $C_s=0$. 

\begin{figure}
\begin{center}
\includegraphics[width=0.3\columnwidth]{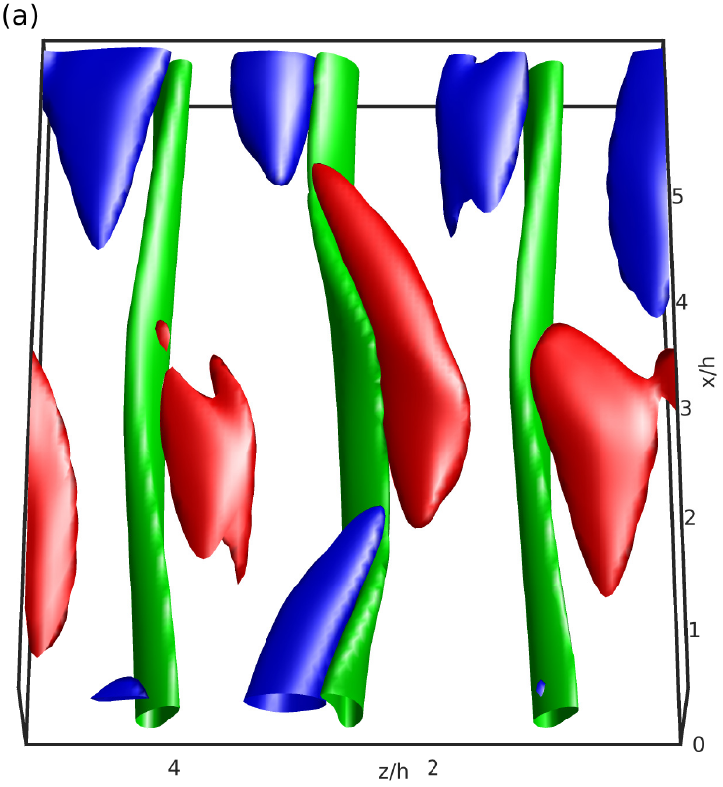}
\includegraphics[width=0.3\columnwidth]{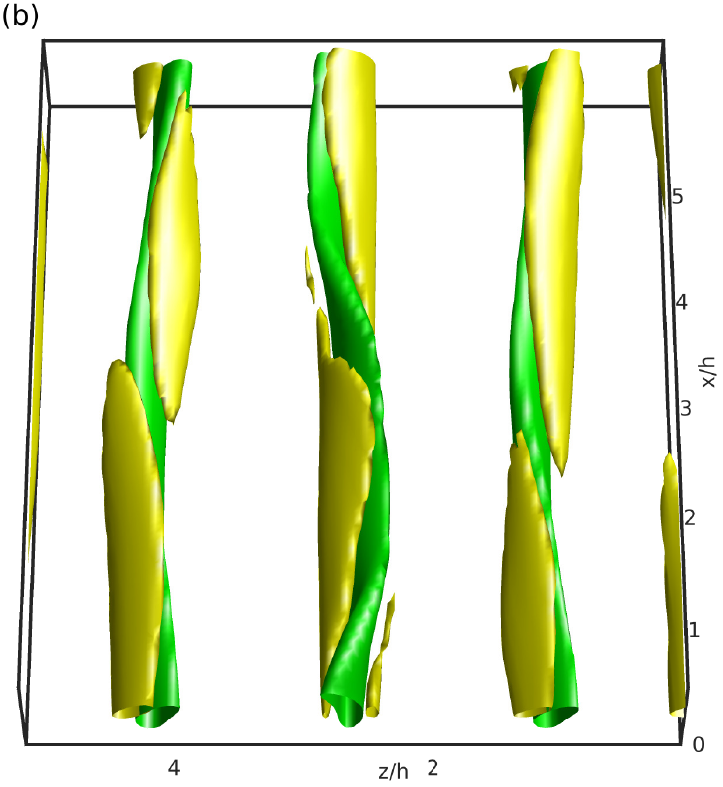}
\end{center}
  \caption{\small  Flow field associated to the upper branch solutions of the filtered (LES) equations  computed at $L_z = 5.55$ for $C_s=0.05$.
The iso-surfaces at $u^+= -2$ are plotted in green. In panel (a) red and blue surfaces correspond to  positive and negative streamwise vorticity at $\omega_x = \pm 0.65\,max(\omega_x)$, while in panel (b) the yellow surface corresponds to $\nu_t/\nu=0.065$.
\label{fig:CS005_UB}
}
\end{figure}

\begin{figure}
 \centering
 \includegraphics[width=0.24 \columnwidth]{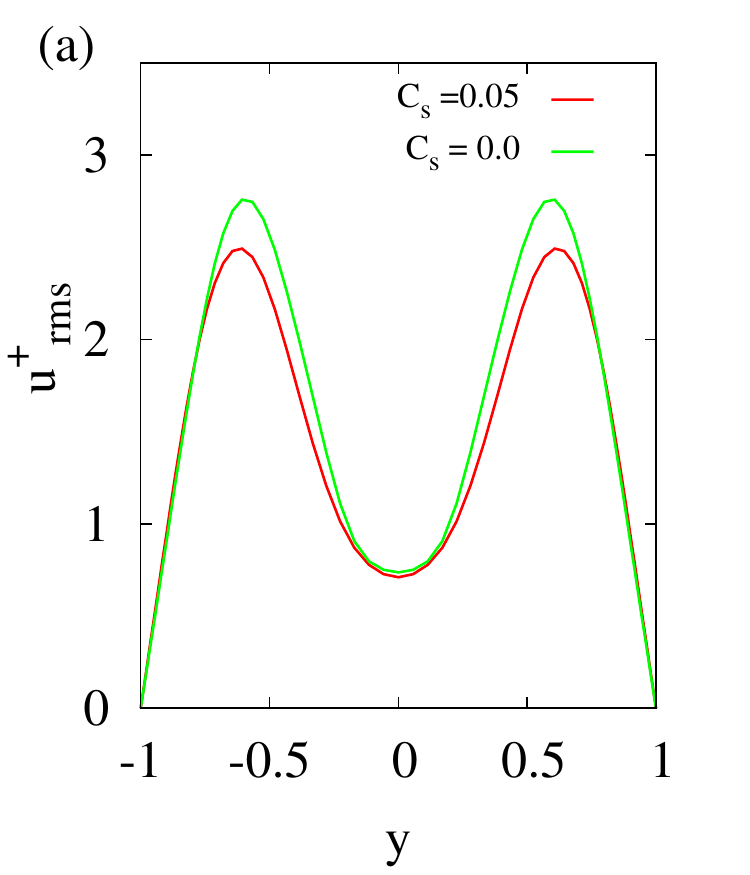}
 \includegraphics[width=0.24 \columnwidth]{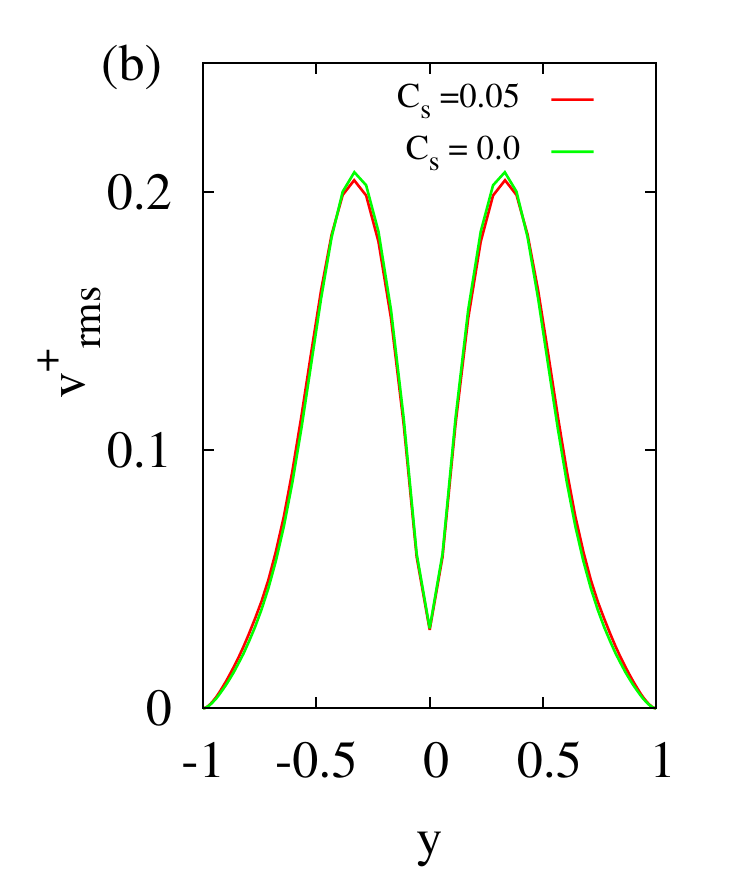}
 \includegraphics[width=0.24 \columnwidth]{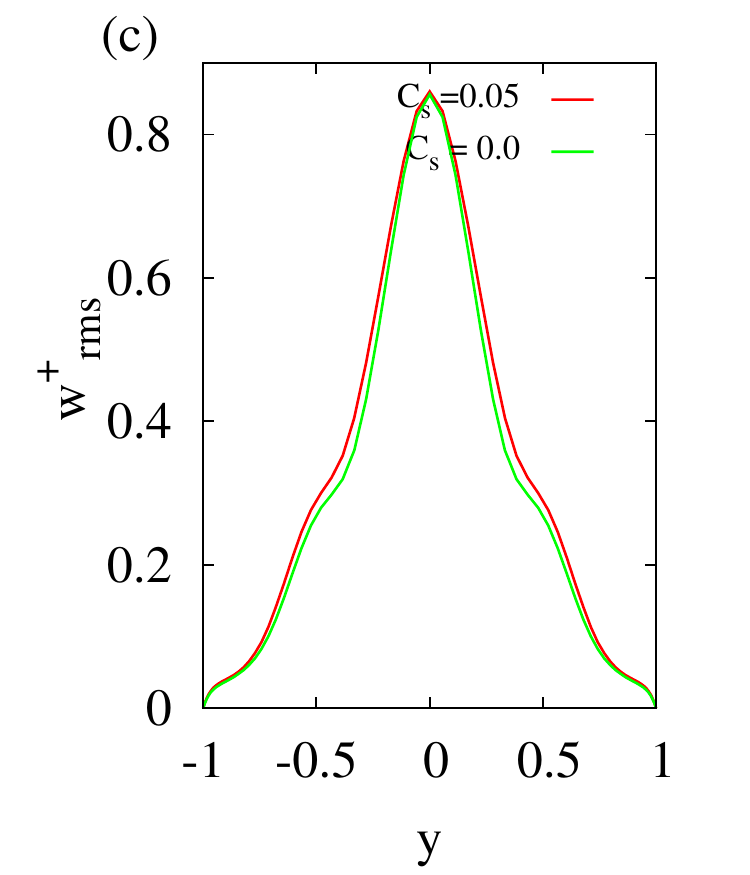}
 \includegraphics[width=0.24 \columnwidth]{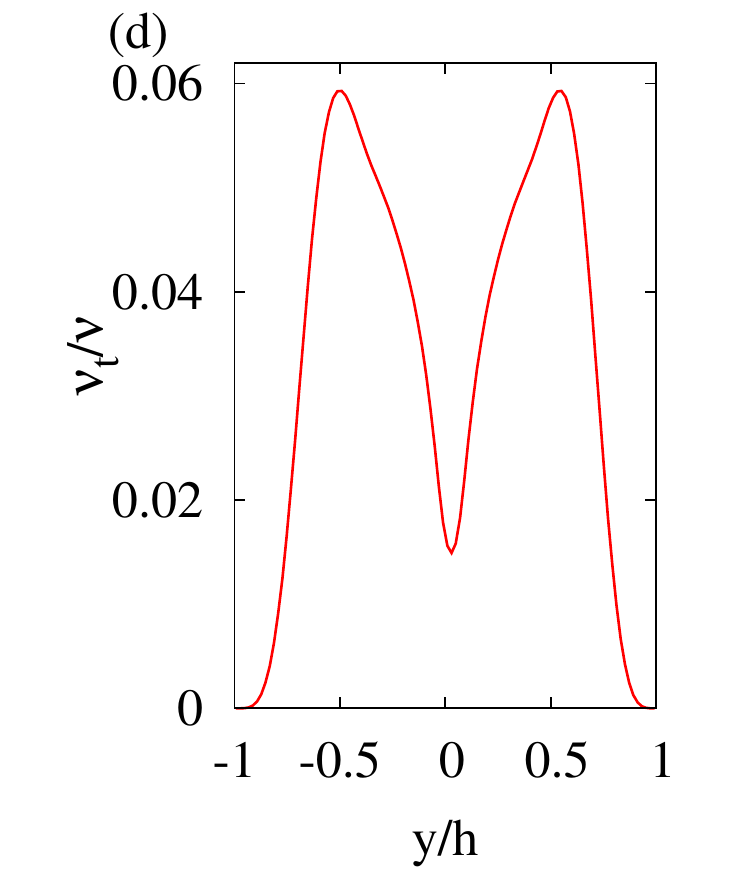}
 \caption{\small Comparison of the $rms$ velocity profiles of the TW upper branch solutions for $C_s=0$ and $C_s=0.05$ (with $L_z=5.55$ and $\Rey=2000$).
In panel (d) is displayed the eddy viscosity $\nu_t/\nu$ associated to the subgrid motions modelled in the solution of the filtered (LES) equations at $C_s=0.05$.
}
 \label{fig:UBLESPOIS}
\end{figure}

We have therefore continued the travelling-wave upper branch solution from $C_s=0$ (Navier-Stokes solution) to $C_s=0.05$ keeping constant the Reynolds number to $\Rey=2000$ and the grid.
This corresponds to taking into account the locally averaged effect of dissipative small scale turbulent motions. 
Proceeding along these lines makes sense because $\Rey=2000$ is more than twice the value of the Reynolds number at which transition is usually observed. 
The continuation proceeds without major problems and convergence is obtained in less than ten steps in $C_s$.
It is found that the introduction of small scale dissipation does not significantly alter the solutions, except for a slight reduction of the streamwise vorticity and of the streamwise velocity, and for the appearance of the sub-grid eddy viscosity shown in figures \ref{fig:CS005_UB} and \ref{fig:UBLESPOIS}.
The computed upper-branch travelling-wave solutions therefore can also be connected to large-scale coherent structures in a fully developed flow. 
To our knowledge, these solutions are the first coherent invariant solutions of the filtered (LES) equations computed for a plane pressure-driven channel.

\section{Discussion and conclusions}

This study has considered invariant solutions in plane Poiseuille flow.
The main results can be summarized as follows:
\begin{itemize}
\item[-] The relative periodic orbit solutions of the Navier-Stokes equations described in Refs.~\cite{Rawat2014} are connected to two branches of travelling-wave solutions via a global saddle-node-infinite period bifurcation when they are continued by increasing the spanwise size $L_z$ of the numerical domain at $\Rey=2000$. 
\item[-] Lower branch travelling-wave solutions remain spanwise localized  when $L_z$ is further increased.
\item[-] Upper branch  travelling-wave solutions develop multiple streaks when $L_z$ is further increased. 
\item[-] Upper branch solutions are not qualitatively changed  when  continued from Navier-Stokes solutions into coherent solutions of the filtered equations (used in large-eddy simulations) by increasing the Smagorisky constant from $C_s=0$ to $C_s=0.05$ as in \cite{Rawat2015b}. 
In the latter case these solutions represent turbulent coherent large-scale motions associated with small-scale turbulent dissipation.
\end{itemize}
Many of these results suggest the existence of some kind of generic dynamics of invariant solutions in wall bounded flows. 
For instance, the dynamics of the relative periodic orbits from where our continuations start (see also \cite{Rawat2014})  is similar to that of relative periodic orbit solutions in the asymptotic boundary layer \cite{Kreilos2013} in their $T/2-L_z/2$ shift property and with their bursting behaviour. 
Also, the saddle-node infinite period bifurcation found by increasing $L_z$ is of the same type of that found in axi-symmetric Rayleigh-B\'enard convection \cite{Tuckerman1988} or in the the bifurcation found by homotopy continuation from Couette flow to the asymptotic suction boundary layer by \cite{Kreilos2013}.
The spanwise localization of the lower branch solution is also in accordance with the results of recent investigations that revealed the spanwise localization of other lower branch solutions \cite{Schneider2010,Duguet2012,Khapko2013,Gibson2014}.

One of the most relevant results shown in this study, however, probably is that the upper branch travelling-wave solutions are seen to develop multiple streaks in the spanwise direction when $L_z$ is increased and that these solutions preserve their structure in the presence of small scale dissipation. 
The wall-normal structure of the upper branch travelling-wave solutions, just as the one of the relative periodic orbits from which they are issued, is reminiscent of large-scale motions in the outer region, even if further work is needed to asses the dynamic relevance of these solutions.
Further investigations are needed to determine if other travelling-wave or relative periodic orbit solutions can be continued to the fully turbulent regime.

\section*{Acknowledgements}
The use of the codes {\tt channelflow} \cite{Gibson2008,Gibson2012}, {\tt diablo} code \cite{Bewley2008} as well as
financial support from PRES Universit\'e de Toulouse and R\'egion Midi-Pyr\'en\'ees are kindly acknowledged.



\newcommand{\noopsort}[1]{} \newcommand{\printfirst}[2]{#1}
  \newcommand{\singleletter}[1]{#1} \newcommand{\switchargs}[2]{#2#1}


\end{document}